# Validation of Continuously Tagged MRI for the Measurement of Dynamic 3D Skeletal Muscle Tissue Deformation

*Kevin M. Moerman [a),b)], Andre M. J. Sprengers [a)], Ciaran K. Simms [b)], Rolf M. Lamerichs [a),c)], Jaap Stoker [a)] and Aart J. Nederveen [a)]*

[a)] Radiology Department, Academic Medical Centre, Meibergdreef 9, 1105 AZ Amsterdam, The Netherlands
[b)] Trinity Centre for Bioengineering, School of Engineering, Parsons Building, Trinity College, Dublin 2, Ireland
[c)] Philips Research, High Tech Campus 11, 5656 AE Eindhoven, The Netherlands

## Abstract

**Purpose:** Typically SPAtial Modulation of the Magnetization (SPAMM) tagged MRI requires many repeated motion cycles limiting the applicability to highly repeatable tissue motions only. This paper describes the validation of a novel SPAMM tagged MRI and post-processing frame work for the measurement of complex and dynamic 3D soft tissue deformation following just 3 motion cycles. Techniques are applied to indentation induced deformation measurement of the upper arm and a silicone gel phantom.

**Methods:** A SPAMM tagged MRI methodology is presented allowing continuous (3.3-3.6 Hz) sampling of 3D dynamic soft tissue deformation using non-segmented 3D acquisitions. The 3D deformation is reconstructed by the combination of 3 mutually orthogonal tagging directions, thus requiring only 3 repeated motion cycles. In addition a fully automatic post-processing framework is presented employing Gabor scale-space and filter-bank analysis for tag extrema segmentation and triangulated surface fitting aided by Gabor filter bank derived surface normals. Deformation is derived following tracking of tag surface triplet triangle intersections. The dynamic deformation measurements were validated using indentation tests (~20 mm deep at 12 mm/s) on a silicone gel soft tissue phantom containing contrasting markers which provide a reference measure of deformation. In addition, the techniques were evaluated *in-vivo* for dynamic skeletal muscle tissue deformation measurement during indentation of the biceps region of the upper arm in a volunteer.

**Results:** For the phantom and volunteer tag point location precision were 44 μm and 92 μm respectively resulting in individual displacements precisions of 61 μm and 91 μm respectively. For both the phantom and volunteer data cumulative displacement measurement accuracy could be evaluated and the difference between initial and final locations showed a mean and standard deviation of 0.44 mm and 0.59 mm for the phantom and 0.40 mm and 0.73 mm for the human data. Finally accuracy of (cumulative) displacement was evaluated using marker tracking in the silicone gel phantom. Differences between true and predicted marker locations showed a mean of 0.35 mm and a standard deviation of 0.63 mm.

**Conclusions:** A novel SPAMM tagged MRI and fully automatic post-processing framework for the measurement of complex 3D dynamic soft tissue deformation following just 3 repeated motion cycles was presented. The techniques demonstrate dynamic measurement of complex 3D soft tissue deformation at sub-voxel accuracy and precision and were validated for 3.3-3.6Hz sampling of deformation speeds up to 12 mm/s.

# I. INTRODUCTION

Since Magnetic Resonance Imaging (MRI) allows fast and 3D imaging with excellent soft tissue contrast without exposing subjects to ionizing radiation, it is an ideal modality for the study of soft tissue motion. The non-invasive analysis of complex soft issue deformations *in-vivo* is relevant to many fields of research such as the study of cardiac biomechanics[1], pre-operative planning[2] and soft tissue mechanical property investigation[3]. The latter is of interest in the current study since the work presented here is part of a study aiming to use inverse analysis of MRI based indentation experiments for the non-invasive evaluation of detailed constitutive models for skeletal muscle tissue. Non-invasive mechanical property analysis is important for a wide range of applications including impact biomechanics[4], rehabilitation engineering[5], tissue engineering[6], surgical simulation[7] and tumor detection[8]. However the evaluation of detailed constitutive models describing the complex mechanical (anisotropic, non-linear and viscoelastic) properties of soft tissues requires detailed experimental deformation measurements capturing the complex (anisotropic, non-linear and time-varying) nature of the soft tissue deformation.

A common approach [1, 9-11] for the MRI-based measurement of soft tissue deformation is to employ SPatial Modulation of the Magnetization (SPAMM) tagged MRI[12, 13]. In SPAMM tagged MRI the tissue is temporarily magnetically tagged using a periodic signal modulation and tracking of the tag pattern allows measurement of deformation.

Typically SPAMM tagged MRI methods require segmented acquisitions whereby an image set reflecting a single motion cycle is composed via repeated imaging of multiple motion cycles. Hence SPAMM tagged MRI has mainly found application in the study of highly repeatable motions such as those occurring in the heart[10]. Recently SPAMM tagged has also been applied to other tissue types in biomechanical studies of repeatedly induced motions of the tongue (e.g. in [14] 16 volunteer speech repetitions per slice), brain (e.g. in [15] 144 volunteer rotational head accelerations) and eyes (e.g. in [16] >135 repeated left to right eye movements). Apart from repeatability constraints, discomfort and health issues may preclude the use of large numbers of repetitions. Especially in the case of biomechanical studies, such as indentations applied in the current study, repeated motions may cause volunteer discomfort. In addition analysis of tissue viscoelasticity and hysteresis is limited. Recently indentation induced SPAMM tagged MRI based 3D skeletal muscle tissue deformation measurement was presented requiring the use of only 3 motion cycles[17]. However the methods were validated only for static deformations (limiting viscoelastic analysis) and the post-processing methods were semi-automatic. Hence the current study focused on the development of dynamic SPAMM tagged MRI based deformation measurement and fully automatic post-processing while maintaining a minimum number of 3 required repetitions.

A wide array of advanced post-processing methods have been proposed for SPAMM tagged MRI (for a detailed discussion see dedicated literature [9, 18] and [19]); for instance using deformable models (see review article [20]), spline models (e.g. [21-24]) and energy based methods such as non-rigid image registration (e.g. [25, 26]) and optical flow methods (e.g. [27]). The post-processing methods in the literature involving deformable models, spline models and non-rigid image registration all inherently require assumptions on the nature of the deformation and/or the mechanical properties and models of the underlying tissue. It is possible to use the above approaches to investigate mechanical properties. However the constitutive model evaluation is limited to the model adopted and difference measures are based on the image data rather than more mechanically meaningful experimental parameters such as 3D deformation, strain and strain rate. Therefore the current study focusses on post-processing methods which allow derivation of the complex 3D deformation while minimizing assumptions on the mechanical properties or the nature of the deformation.

Most of the above post-processing methods focus on magnitude image data. However specialized methods have also been developed for tag phase analysis. This includes HARmonic Phase (HARP) (e.g.[28]). In HARP tag spectral peaks are isolated through the application of band-pass filtering, and inverse Fourier transformed yielding a tag free magnitude image and an approximate tag phase image. Under the assumption of phase invariance (i.e. that material points maintain the same "harmonic" phase under deformation) phase can be viewed as a material property and used for motion estimation. However, the derivation of deformation often requires phase unwrapping which is an error prone process. The design of an appropriate band-pass filter is not trivial and may influence the deformation information obtained (i.e. the smaller the selected peak region the straighter the obtained line pattern) and since a global frequency of tags is assumed local deformations may not be accounted for appropriately. In addition the appropriate separation of the tag spectral peaks is hindered by the spectral power associated with the DC peak. The influence of the DC peak can be reduced through techniques such as Complimentary SPAMM (CSPAMM) [29] or the more recently proposed Total Removal of Unwanted Harmonic Peaks TruHARP [30]. However these techniques require the use of additional repetitions and are thus not of interest to the current study.

In an effort to reduce some of the issues associated with HARP analysis (e.g. phase inconsistency induced bifurcations) Qian *et al.* 2003 [31] introduced Gabor filter banks for SPAMM tagged MRI analysis. A Gabor filter [32] is a Gaussian function modulated by a (possibly complex) harmonic function. Since Gabor filters are also periodically modulated structures they are ideally suited for SPAMM tagged MRI analysis and are increasingly applied for this purpose[33-43]. In Gabor filter analysis an array or bank of filters is applied each with different characteristics (e.g. harmonic frequency, orientation, phase and Gaussian envelope size). The final Gabor filter bank response is then composed from a combination of the various filter responses e.g. using the maximum filter response for each voxel. Gabor filters are most often used for frequency domain and phase analysis and hence in some cases serve as an alternative band-pass filter for HARP type analysis (e.g. [36, 40]) which following phase unwrapping enables derivation of displacement estimates. However in many cases the error prone phase unwrapping is avoided and analysis focuses on tracking of the wrapped phase data in which a saw-tooth tag phase modulation is present (e.g. [33]). Mostly 2D Gabor filters have been employed (e.g. [33, 35, 40, 41]) allowing for derivation of 2D displacement estimates. Such 2D measures have also been used to drive 3D deformable models for 3D deformation estimation (e.g. [35]). Recently 3D Gabor filter banks have been employed (e.g. [38, 39]) these have focused not on reconstructing phase data but on filter optimized magnitude image data. For instance Shimizu *et al.* 2010 [38] used a 3D Gabor filter bank to enhance tag appearance. Tags were segmented as local minima and tracking of tag surface triplet intersection points, determined using iterative minimization, provided 3D displacement. These methods were however semi-automatic and employed a smooth energy minimizing Snakes function for tag segmentation which regularizes the deformation and limits applicability to complex deformations. This review shows that Gabor filters may be valuable aids in the analysis of SPAMM tagged MRI data but that automatic post-processing methods for 3D deformation analysis have not been proposed yet without the need for regularization structures such as splines functions or deformable models.

Gabor filters themselves have a regularizing effect through blurring or smoothening of the image data. Since the smoothening suppresses noise and local disturbances it aids in segmentation of tag features. However the smoothening effect also suppresses local deformation information. The amount of smoothening depends on the size of the Gaussian envelope which confines the filter. Typically Gaussian standard deviations in the order of the tag (or Gabor) period (1/tag or Gabor frequency) are used (e.g. [36, 39, 40]). Such standard deviations lead to significant blurring across a region more than 5 times the tag period in width. Thus although this provides improved tag feature segmentation the amount of local deformation information obtainable is reduced (as large standard deviations straighten the tag features). Smaller standard deviations do not over regularize deformation but contain more features which may be detrimental to segmentation. In the current study a novel approach is presented whereby the best of both types of analysis (large and small Gaussian standard deviations) is combined. Besides a Gabor filter bank, a Gabor scale-space (a non-linear scale-space where "scale" corresponds to Gaussian standard deviation magnitude) is proposed which aids in segmentation using large standard deviations, however the segmentation is adjusted to filtered image data of increasingly smaller standard deviations rendering a final segmentation result which contains a maximum of deformation information.

Gabor filters are very versatile since besides tag enhancement they may also serve as direct "measuring tools" of image features which relate to the Gabor parameters independent of segmentation. This is because the Gabor filter produces the maximum response when local image features resemble its appearance. For instance the Gabor frequency which yielded the maximum response relates to the local (within the Gaussian envelope) tag spacing and may be used to directly estimate strain measures (e.g. as proposed and demonstrated in a numerical phantom in [43]). Similarly the Gabor filter orientation which produced the maximum response relates to the local tag orientation. The latter is implemented in the current study as an aid in surface fitting. Following segmentation of tag features some studies employ continuous and smoothening surface fits to represent tags (e.g. spline functions such as in [17]). These generally inaccurately represent complex deformations such as sharp transitions and separated tag features (e.g. induced by shear separation and sliding interfaces). In the current study tag surfaces are fitted through orthogonal weighted averaging of the segmented tag features, whereby the orthogonal directions for fitting are derived from the Gabor orientation estimates. The surface connectivity is created via Delaunay triangulation incorporating gaps and separations. Surfaces are then mildly smoothened to suppress noise induced fluctuations however smoothening is local and does not occur across separations or gaps.

This paper presents a novel (non-segmented) SPAMM tagged MRI methods enabling continuous sampling of complex 3D dynamic tissue deformation using 3 motion cycles only. The post-processing framework is computationally efficient and fully automatic and features: Gabor filter bank and scale-space analysis, scale-space assisted tag segmentation, orthogonal weighted averaging and Delaunay triangulation based surface fitting. Deformation is measured following tracking of mutually orthogonal surface triplet intersections. Since no regularizing (deformable/spline) model is required for computation of deformation, the post-processing framework presented avoids many assumptions on the tissue deformation and is thus ideal for the analysis of complex deformation (i.e. involving tissue non-linearity, anisotropy and sliding induced shearing). For validation the techniques are applied to the measurement of complex dynamic deformation in a silicone gel phantom containing markers which, when tracked, provide a reference measure of deformation enabling derivation of technique accuracy[44]. In addition the techniques are applied to indentation induced deformation measurement in the upper arm of a volunteer for *in-vivo* evaluation.

## II. METHODS

## II.A. The experimental set-up: indentor and soft tissue phantom

In the current study 3D dynamic deformation is measured using continuously sampled dynamic SPAMM tagged MRI (section II.B) applied in 3 orthogonal directions during 3 motion cycles. In order to validate the deformation measurements performance was evaluated in a silicone gel tissue phantom and *in-vivo* in the upper arm of a volunteer (FIG. 1). A computer controlled MRI compatible indentor with a flat circular (40 mm in diameter) head was used to apply repeated transverse indentation (~20 mm deep) to a silicone gel phantom and the biceps region of the upper arm of a volunteer. Since 3 repeated motion cycles are required for derivation of 3D deformation the indentor motion was triggered (using a scanner generated TTL pulse) to start after the first dynamic of each acquisition series. This first dynamic thus provides the initial un-deformed tag pattern state. A single motion cycle is defined as an indentation phase, a hold phase and a retraction phase. For validation of the deformation measurement the silicone gel soft tissue phantom (200 mm long and 120 mm in diameter and a stiff 20 mm in diameter bone-like core) contains contrasting (low signal) spherical markers (3±0.05 mm in diameter). These markers were tracked[44] from T2-weighted scans (0.5 mm isotropic) of the same field of view for the initial un-deformed and the final deformed configuration. In addition phantom and *in-vivo* accuracy measures could be derived from the fact that the displacement paths should return to their original location as displacement is recorded up to the end of the retraction phase of the motion cycle. For more information on the experimental set-up the reader is referred to [17, 44].

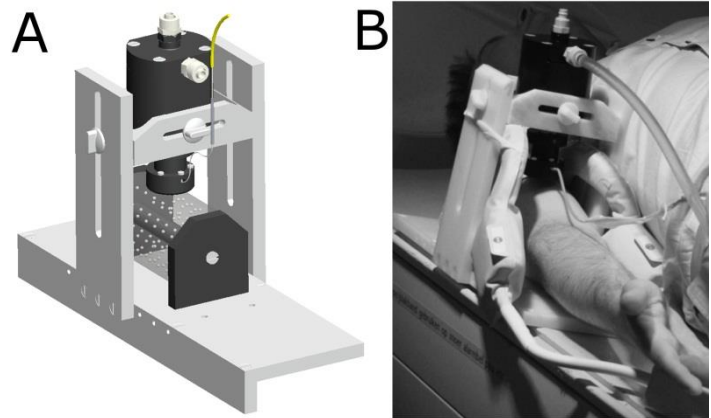

FIG. 1 The experimental set-up showing the MRI compatible indentor used for indentation of a silicone gel phantom (A) and volunteer upper arm (B).

## II.B. MRI sequence

In the current study a (non-ECG triggered) single shot SPAMM tagged MRI sequence was employed which is a dynamically optimized version (read-out acceleration and delay reduction) of the sequence presented in[17]. A schematic for the pulse sequence design is shown in FIG. 2. TABLE I provides a summary of the scanning parameters and configurations used in this study and illustrates several slices for each data set recorded. As the sequence diagram shows, following a 1-1 SPAMM tag pre-pulse (parts A) a short delay (parts B) was introduced during which the tissue and tag pattern deforms. Subsequently the image data is acquired using a single 3D Transient Field Echo (TFE) read-out (parts C). The 3D TFE read-out was configured with a Cartesian acquisition mode in k-space, the profile order was set to low-high, a radial turbo direction was used and half-Fourier was applied for acceleration ($0.625$ in the readout direction and 0.8 in the first phase encode direction). All scans were performed on a 3.0T (Intera, Philips Health Care, Best, The Netherlands) MRI scanner using flexible surface coils with two elliptical elements (diameters of 14 and 17 cm) placed laterally to the upper arm (FIG. 1B). Each individual acquisition was non-segmented and not repeated and entire image volumes were acquired consecutively in time for each direction. For the phantom and volunteer imaging the motion was thus effectively continuously sampled at 3.3 Hz (123 ms delay + 177 ms read-out) and 3.6 Hz (100 ms delay + 177 ms read-out) respectively. Full 3D dynamic deformation measurement is achieved through the combination of dynamic SPAMM data from 3 orthogonal directions, thus requiring 3 motion cycles. However for validation purposes multiple motion cycles (see TABLE I) were recorded for each acquired SPAMM direction allowing analysis of technique accuracy and precision both in the phantom and *in-vivo*. Thus for the phantom and volunteer tests 11 and 37 consecutive dynamics per motion cycle were acquired respectively (5 during indentation phase 1 during hold phase and 5 during retraction phase for the phantom, and 9 during indentation phase, 15 during hold phase and 13 during retraction phase for the volunteer). For the volunteer indentation was applied at a lower speed to ensure comfort and with a longer hold phase to allow for viscoelastic recovery. Although the overall deformation magnitudes are similar to those tested in[17] the individual deformations for each dynamic were relatively low (up to the tag period in magnitude) since the deformation (occurring at ~12 mm/s) was continuously sampled at 3.3-3.6Hz.

During each dynamic acquisition motion is allowed to continue during the read-out. Therefore temporal blurring may cause mild underestimation of the motion. The amount of underestimation is related to many factors including the nature of the read-out and the deformation (speed and directions). The spatial characteristics of SPAMM tags correspond with specific peaks in the k-space domain, which define the bulk of the deformation information encoded in the tag pattern. However frequency components from the whole of the k-space domain contribute to the more detailed local deformation information encoded in the tag pattern. Therefore, since the read-out profile order was low-high, the latter is acquired towards to the end of the read-out sequence. As such it is expected that significant motion features are still acquired towards the end of the read-out and no compensation of the possible underestimation is required.

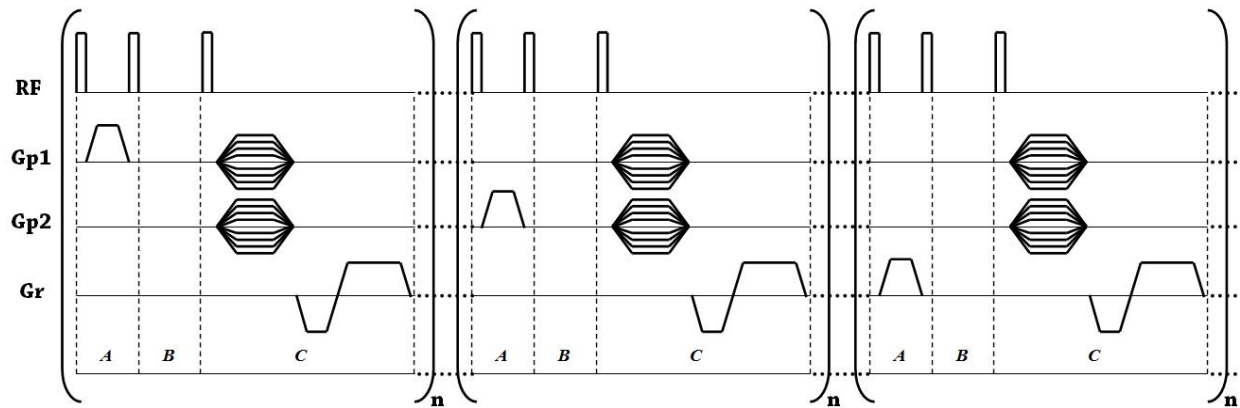

FIG. 2 Diagram of the SPAMM pulse sequence and 3D TFE read-out. The 1-1 SPAMM pre-pulse (A) modulates the signal followed by a desired time delay (B) after which a 3D volume is acquired (C). The acquisition can be repeated n times for each direction.

TABLE I The MRI acquisition matrix

| MRI data visualisations: Iso-surface of indented state, and a selection of 5 evenly spaced slices for each image data set | Scan type # dynamics, #indentations $T_R/T_E$ (ms) Read-out time (ms) | Field of view (mm), Acquisition matrix, # slices Reconstructed voxel size (mm) | Slice orientation, Tag parameters: $p_t$ (mm), $\theta_t$ (°), $\psi_t$ (°) Delay time (ms) |
|---|---|---|---|
| **Phantom imaging** | | | |
| 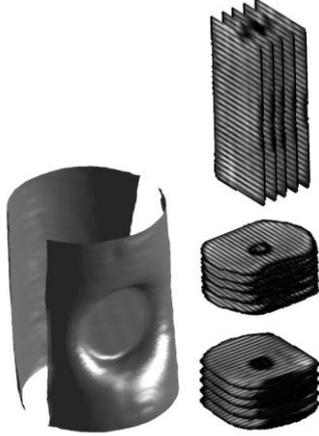 | 1-1 SPAMM<br>12, 1<br>2.39/1.16<br>177 | 120x120x39<br>80x80, 13<br>0.93x0.93x1.5 | Sagittal<br>4, $0.5\pi$, 0<br>123 |
| | 1-1 SPAMM<br>48, 4<br>2.38/1.53<br>177 | 120x120x39<br>80x52, 13<br>0.93x0.93x1.5 | Transversal<br>4, 0, 0<br>123 |
| | 1-1 SPAMM<br>48, 4<br>2.53/1.28<br>177 | 120x120x39<br>80x52, 13<br>0.93x0.93x1.5 | Transversal<br>4, $0.5\pi$, 0<br>123 |
| | T2-weighted<br>2, 1<br>2500/638<br>10min | 120x120x80<br>240x240, 160<br>0.47x0.47x0.5 | Sagittal<br>N.A.<br>N.A. |
| **Volunteer imaging** | | | |
| 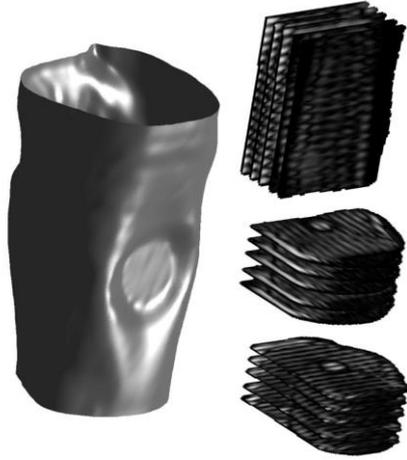 | 1-1 SPAMM<br>150, 4<br>2.45/1.21<br>177 | 100x100x40<br>68x52, 10<br>0.89x0.89x2 | Coronal<br>6, $0.5\pi$, 0<br>100 |
| | 1-1 SPAMM<br>150, 4<br>2.42/1.19<br>177 | 120x120x40<br>80x60, 10<br>0.94x0.94x2 | Transversal<br>6, 0, 0<br>100 |
| | 1-1 SPAMM<br>150, 4<br>2.57/1.30<br>177 | 100x100x40<br>68x52, 10<br>0.89x0.89x2 | Transversal<br>6, $0.5\pi$, 0<br>100 |

## II.C. Spatial characteristics of SPAMM tags

To aid the description of the analysis methods in section II.D the spatial characteristics of SPAMM tags are first briefly discussed. For a more detailed discussion on SPAMM tagged MRI the reader is referred to[10]. SPAMM tagging induces a periodic modulation on the signal profile across the image volume and in the case of 1-1 (first order) SPAMM is approximately sinusoidal. Thus an un-deformed tagged image $M_t$ can roughly be expressed as the following type of modulation of a magnitude image $M$:

$$M_t \approx \left| M \cdot \left( m + A \cdot cos\left(\frac{2\pi x_r}{p_t}\right) \right) \right| \qquad \text{II-1}$$

Here $p_t$, $m$ and $A$ set the spatial tag period (or spacing), mean and amplitude of the modulation respectively across the direction $x_r$. Throughout this paper tag modulations are expressed along $x_r$ belonging to the coordinates $(x_r, y_r, z_r)$ formed following rotation of $(x, y, z)$, expressed in a regular Cartesian coordinate system:

$$(x_r, y_r, z_r) = R_t \cdot (x, y, z) \qquad \text{II-2}$$

Where $R_t$ represents the rotation matrix:

$$R_t = \begin{bmatrix} cos(\theta_t) & 0 & cos(\theta_t) \\ 0 & 1 & 0 \\ -sin(\theta_t) & 0 & sin(\theta_t) \end{bmatrix} \begin{bmatrix} cos(\psi_t) & -sin(\psi_t) & 0 \\ sin(\psi_t) & cos(\psi_t) & 0 \\ 0 & 0 & 1 \end{bmatrix} \qquad \text{II-3}$$

expressed using the Euler angles $(\theta_t, \psi_t)$ which define an x-axis and z-axis rotation consecutively. The initial and un-deformed tag orientation and period are set during scanning. In the current study 3 mutually orthogonal SPAMM directions were applied and their orientation and frequency characteristics in the above notation are presented in TABLE I.

For each image data set a local Cartesian coordinate system $(x, y, z)$ is defined which is aligned with its image axes (voxel row, column and slice directions). The field-of-view location and orientation which are set during scanning determine the local coordinate system's origin location and axes orientations. Local coordinate systems thus vary for each image data set and generally do not coincide with the overall scanner coordinate system (based on bore axis and perpendicular directions). The tag orientation parameters $(\theta_t, \psi_t)$ are defined with respect to the local (not overall scanner) coordinate system. Thus for two image sets with orthogonal tag features in space (the overall scanner coordinate system) the respective tag orientation parameters $(\theta_t, \psi_t)$ may be equivalent (TABLE I) as long as the set field-of-view orientations are mutually orthogonal. Although throughout the work presented here computations are mainly performed in the overall scanner coordinate system the methods and results are presented in local coordinates systems for clarity as features may be oblique with respect to the global scanner coordinate system.

Although an un-deformed tagged image will have the features discussed above, tissue deformation causes them to vary spatially depending on the nature of the deformation (e.g. local orthogonal tension or compression will increase or decrease $p_t$ respectively and bending will locally perturb $\theta_t$ and $\psi_t$). However this can be taken into account during analysis and the periodic nature will remain advantageous in segmentation as will be discussed in section II.D.1. In addition field inhomogeneity and local material property differences may cause the tag features to be non-planar prior to the onset of deformation. As such for the current study dynamics acquired in the absence of motion were used as the reference state.

The extrema (maxima and minima) of the periodic modulation are here referred to as tags or tag features and surfaces fitted to them as tag surfaces. An intersection point for a tag surface triplet is referred to as a tag point. In order to derive deformation the current study applies segmentation of the extrema and tracking of mutually orthogonal tag feature derived tag points. This leads to a trackable grid of tissue points with a $\frac{p_t}{2}$ mm spacing (2x2x2 mm and 3x3x3 mm for the phantom and volunteer data respectively).

## II.D. Deriving tissue deformation from the SPAMM tagged MRI data

In order to derive full 3D dynamic soft tissue deformation from the continuous dynamic SPAMM tagged MRI data a post-processing framework was created and implemented using MATLAB (7.8 R2009a The Mathworks Inc., Natick, MA). Magnitude MRI data was imported and normalized for each dynamic and post-processing was performed in the following 5 steps:

*1) Gabor filter analysis:* The goal of this analysis step is *a)* to derive (filtered) image data sets which will aid in the successful segmentation of tags, and *b)* to derive per voxel tag surface normal orientations which will aid the surface fitting methods presented.

*2) Tag feature segmentation:* The goal of this analysis step is to segment and separately group tags based on the Gabor image data sets.

*3) Orthogonal weighted-mean surface fitting:* Using the per voxel tag orientations estimated from the Gabor filter bank a triangulated surface description is created where surface points are derived using orthogonal weighted means.

*4) Tag surface intersection determination*: Using the triangulated surface descriptions tag intersection points from the 3 mutually orthogonal directions are derived producing the tissue points trackable over time.

*5) Derivation of displacement fields*: This analysis step produces the dynamic and cumulative displacement fields.

### *II.D.1. Gabor Filter Analysis*

Due to the specific spatial characteristics of SPAMM tags (discussed in section II.C) they can be enhanced using a filter which shares these characteristics, the Gabor filter[32] (see FIG. 3A-B). A Gabor filter is a wavelet constructed by modulating a Gaussian envelope with a harmonic function (FIG. 3A and B visualize a 1D and 3D example). For the current study magnitude SPAMM tagged MRI data is used and is analyzed using the following type of 3D Gabor filter:

$$G(x_r, y_r, z_r, \sigma_p, \sigma_s, p_g) = e^{-\frac{1}{2}\left(\left(\frac{x_r}{\sigma_p}\right)^2 + \left(\frac{y_r}{\sigma_s}\right)^2 + \left(\frac{z_r}{\sigma_s}\right)^2\right)} \cdot \left(\pm cos\left(\frac{2\pi x_r}{p_g}\right)\right) \qquad \text{II-4}$$

Here $p_g$ is the central period of the harmonic modulation and the parameters $\sigma_p$ and $\sigma_s$ define the size of the Gabor filter since they represent the perpendicular, and within-tag surface standard deviations respectively for the ellipsoidal Gaussian envelope. The $\pm$ denotes alteration of sign when tracking of the maxima (+) or minima (-) is of interest. Analogous to the equations and notation introduced in section II.C the modulation acts along the $x_r$ direction in the coordinate system $(x_r, y_r, z_r)$ formed following rotation of the system $(x, y, z)$ with a rotation matrix defined using the Gabor angles $(\theta_g, \psi_g)$. Given its particular frequency, size and orientation, convolution with the Gabor filter will produce an image where features that locally resemble its appearance are amplified while others are suppressed. However, as discussed in section II.C, motion and deformation results in locally varying tag frequency and orientation. Therefore a common approach[33, 35, 38, 39] is to employ an array or bank of filters, all with different spatial and frequency characteristics. A single filtered image can then be reconstructed by taking the maximum response of all filters for each voxel. For computational efficiency all convolutions were computed as:

$$M_f = \mathcal{F}^{-1}\{\mathcal{F}\{M_t\} \cdot \mathcal{F}\{G\}\} \qquad \text{II-5}$$

Here $M_f$ represents the filtered image and $\mathcal{F}\{\ \}$ and $\mathcal{F}^{-1}\{\ \}$ denote the Fourier and inverse Fourier transform respectively.

It is important to note that the choice of the Gabor parameters can affect the apparent deformation derived. For instance the standard deviations set the amount of blurring in their respective directions which might cause undesired averaging effects on the deformation (large standard deviations straighten the tag features). Therefore for the filter bank used $\sigma_p$ was set at $\frac{p_g}{3}$. Hence when centred on an extreme, in the direction orthogonal to the modulation the filters only act on the extrema centre and its two directly neighbouring extrema. For a visualisation of a similarly confined 1D Gabor wavelet see FIG. 3A. The within-tag-surface standard deviation $\sigma_s$ sets the amount of averaging along the surface (e.g. size of disks in FIG. 3B) and large values will have a straightening effect on the tag shape while smaller values leave its original shape and detail intact. Using large standard deviations effectively has a smoothening and regularisation effect not only on the images, but also on the derived deformation. This however is desirable for segmentation as the effects of noise are suppressed allowing easy separation of more blur-scale invariant features such as tags. To harness this benefit without over-regularizing the deformation the Gabor analysis presented here is split into two parts. Besides a Gabor filter bank, a Gabor scale-space (a non-linear scale-space where "scale" refers to blur level or magnitude of $\sigma_s$ and $\sigma_p$) is proposed which aids in segmentation (see section II.D.2) using large $\sigma_s$ and $\sigma_p$ values. However the segmentation derived for the high $\sigma_s$ and $\sigma_p$ values is iteratively adjusted to filtered image data of increasingly smaller $\sigma_s$ and $\sigma_p$ values. Following adjustment of the segmentation to the last least blurred image (with the smallest $\sigma_s$ and $\sigma_p$ values) the segmentation is adjusted to a Gabor filter bank optimised image data rendering a final segmentation result where deformation has not been over-regularized.

The Gabor scale-space and Gabor filter bank filtering are schematically illustrated in FIG. 4 and the parameters used are specified in TABLE II. For the Gabor scale-space the orientation and central period were held constant while for both the perpendicular and within-surface standard deviations a scale-path or range was specified going from $p_t$ to $\frac{p_t}{3}$ in 6 scale steps (which translates to blurring in a region approximately 5 tag periods wide down to blurring up to just the neighbouring extrema). For each scale-space filter a separate image was formulated leading to 6 scale-space image sets (e.g. FIG. 5B and FIG. 5C represent two different scale steps) for each extrema type.

For the Gabor filter bank the filter orientation and central frequency were varied across 27 filter combinations. The within-surface standard deviation was held constant while the perpendicular standard deviation was constrained as $\frac{p_g}{3}$ such that the filter design of FIG. 3A is maintained. The maximum filter response (in the image domain) was used for each voxel producing two filtered image sets, one for each extrema (e.g. FIG. 5D). Besides aiding in segmentation (see section II.D.2) the Gabor filter bank was also used as a "measuring tool" since the specific Gabor filter orientation that produced the maximum response was also stored for each voxel as this is an estimate of the local tag surface orientation and can be used in surface fitting (see section II.D.3). Specifically it provides an estimate of the local surface normal vector $\mathbf{n_s}$ since:

$$\mathbf{n_s} = \begin{bmatrix} \cos(\theta_g)\cos(\psi_g) \\ \sin(\psi_g) \\ -\sin(\theta_g)\cos(\psi_g) \end{bmatrix} \qquad \text{II-6}$$

The surface normal estimates are only appropriate for locations that actually resemble the filter such as the central voxels for the tag extrema. The surface normal orientations are thus only used for these voxels (see also II.D.3).

For each extrema type the Gabor filter analysis produced: 6 Gabor scale-space image sets (going from most to least blurred in 6 steps), a Gabor filter-bank optimized image set, and a surface normal estimation data set.

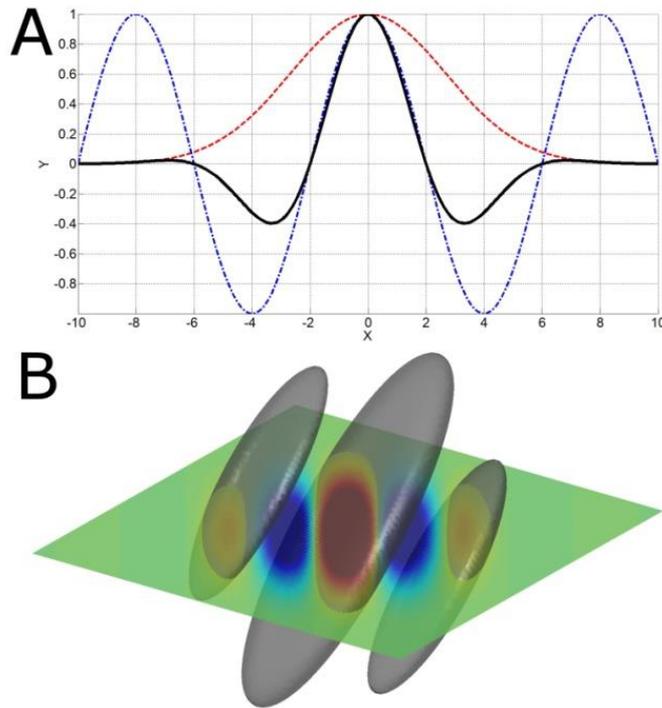

FIG. 3 1D representation of a Gabor wavelet (solid) composed using multiplication of Gaussian (dashed) with harmonic function (dashed-dotted) (A), 3D visualisation of a Gabor wavelet showing 2D mid-slice and iso-surfaces (B).

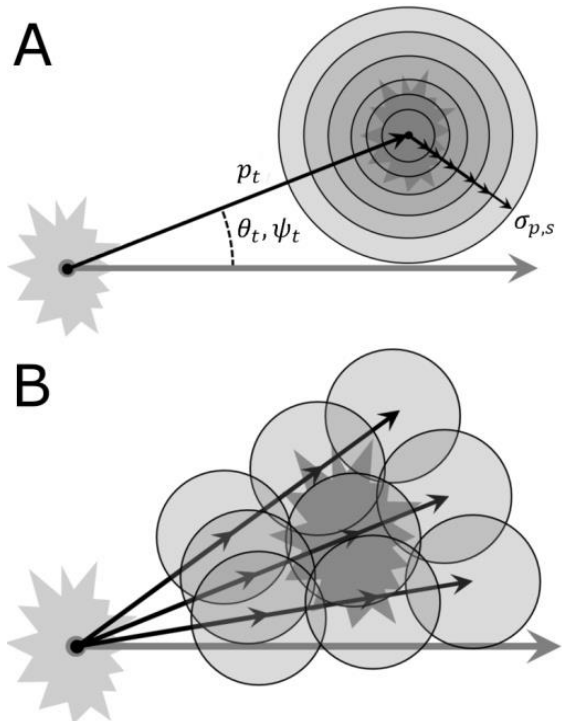

FIG. 4 Schematic representation of the Gabor scale space (A) and Gabor filter bank (B). Circle centre distance from origin, circle radius and location relate to the Gabor central period, Gaussian envelope size and orientation respectively.

TABLE II The Gabor filter designs

| Gabor filter set type | x-axis tilt $\theta_g$ (°) | z-axis tilt $\psi_g$ (°) | Central period $p_g$ (mm) | Perpendicular standard deviation $\sigma_p$ (mm) | Within surface standard deviation $\sigma_s$ (mm) |
|---|---|---|---|---|---|
| Gabor scale space filters (6 scale steps) | $\theta_t$ | $\psi_t$ | $p_t$ | $p_t \to \dfrac{p_t}{3}$ | $p_t \to \dfrac{p_t}{3}$ |
| Gabor filter bank (all 27 combinations) | 3 variations: $\theta_t - 20$ $\theta_t$ $\theta_t + 20$ | 3 variations: $\psi_t - 20$ $\psi_t$ $\psi_t + 20$ | 3 variations: $p_t - 1$ $p_t$ $p_t + 1$ | $\dfrac{p_g}{3}$ | $\dfrac{p_t}{3}$ |

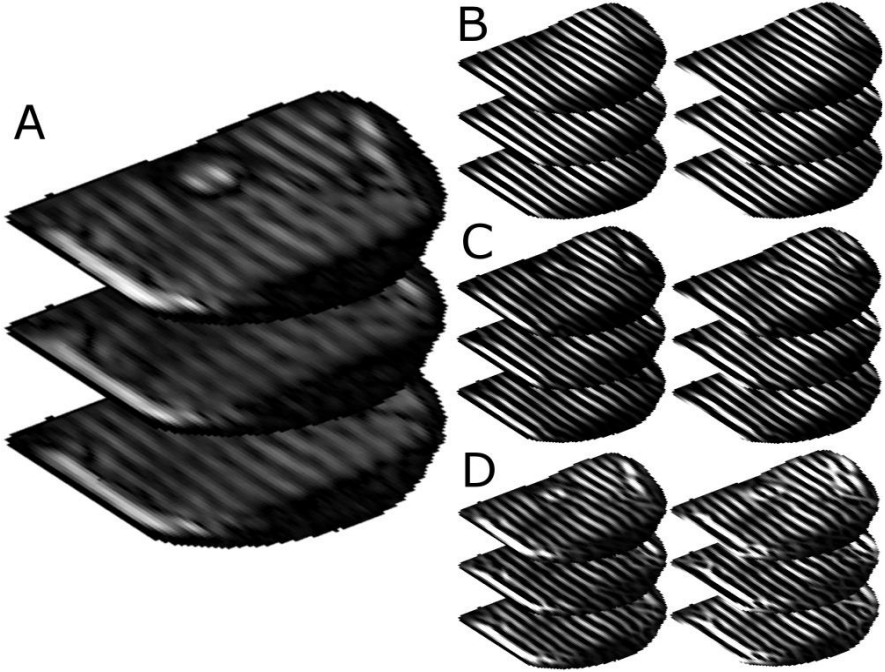

FIG. 5 An unfiltered image (A), and Gabor scale-space filtered images for the most blurred step (B) and an intermediate step (C) and Gabor filter-bank enhanced images (D). Maxima are shown on the left and minima on the right in (B-C). The scale-space images show separate and straightened tag features while the filter-bank images

### II.D.2. Tag feature segmentation

The first step in tag segmentation is to produce logic images reflecting whether voxels in all the Gabor-filtered image sets are potential tag voxels. An appropriate and adaptive threshold $T_{tag}$ was half the mean of all voxels within a set with an intensity higher than 0 (i.e. all significantly positively enhanced voxels, since features dissimilar or opposite in nature to the Gabor filter become reduced or negative). The logic images $L_{tag}$ for each Gabor image set $M_f$ were derived using:

$$L_{tag} = \begin{cases} 1 & M_f \geq T_{tag} \\ 0 & M_f < T_{tag} \end{cases} \quad T_{tag} = \frac{\sum M_h}{2 \sum L_h} \quad M_h = \begin{cases} M_f & M_f > 0 \\ 0 & M_f \leq 0 \end{cases} \quad L_h = \begin{cases} 1 & M_f > 0 \\ 0 & M_f \leq 0 \end{cases} \quad \text{II-7}$$

Then a simple grouping algorithm was implemented whereby all potential voxels that are touching each other with one of their 6 faces are grouped together to form a tag feature. The initial tag features are formulated by grouping in the most blurred Gabor filtered image (FIG. 5B and FIG. 6A) since here all tag features are appropriately separated. As FIG. 6 demonstrates the threshold $T_{tag}$ segments the tag features with a thickness of about $\frac{p_t}{2}$. Next the shape of each tag feature is adjusted using the following step wise process repeated for each less blurred Gabor scale-space image (6 steps) and finally also the Gabor filter-bank optimised image (ensuring tag spacing and orientation variations are also appropriately accounted for) and thus a total of 7 tag feature adjustment steps are used:

1) An orthogonal weighted average (perpendicular to tag feature orientation) of the voxel coordinates of the current tag feature shape is taken to provide reference coordinates of the voxels at the centre of the tag (leaving the tag feature 1 voxel thick).
2) The tag feature shape for the next image set is then defined as all voxels that are classified as potential tag voxels and are touching one of the central tag voxels from the previous step (this regrows the tag features to their normal $\frac{p_t}{2}$ thickness, this step can be repeated for tags with large $p_t$ relative to the voxel size).

After the final step the tag shape has also been adjusted to the Gabor filter bank enhanced image which maintains features of deformation (FIG. 5D). Segmentation is performed for each dynamic producing results similar to FIG. 6. During segmentation each tag feature is also assigned with a tag number enabling indexing of tag intersection points (section II.D.4). First the tag features in the initial configuration are numbered. Tag features in consecutive dynamics are then numbered by finding the closest (within a distance < $p_t$) corresponding tag in the initial configuration. This approach is enabled since tag deformations are relatively small due to the fast imaging employed in the current study with respect to the applied deformation. The mapping to the initial configuration ensures that possible missing or additional tags are appropriately accounted for.

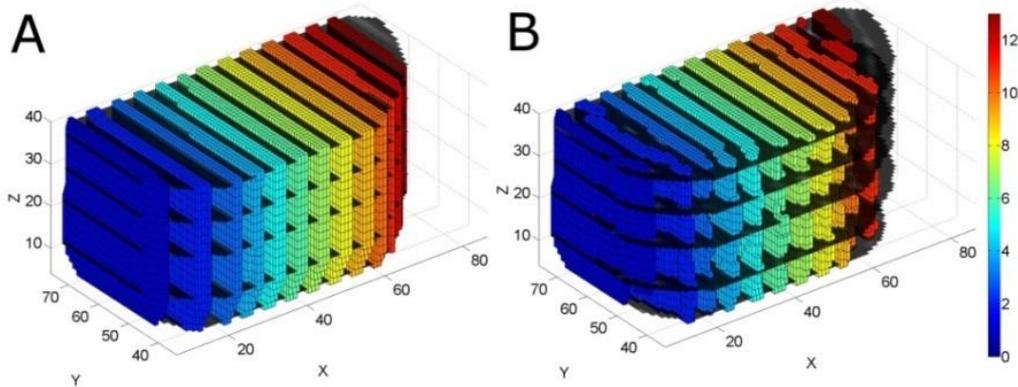

FIG. 6 Segmented tag maxima features shown as voxels (shaded according to tag number) in the most blurred (A) and final Gabor bank filtered state (B). Axis units are in mm.

*II.D.3. Orthogonal weighted-mean surface fitting*

A simple yet flexible surface fitting approach was derived as it can fit surfaces of any orientation and severe deformation and voids due to shearing or gaps. First central tag feature voxels are identified. Most central tag voxels can easily be identified as tag voxels which present with orthogonal neighbors (as tag voxels at the periphery of a tag do not). However thin tag regions (e.g. less than 3 voxels thick), which present with insufficient orthogonal neighbors, require special treatment. Here central tag voxels could be uniquely identified as all tag voxels which do not have orthogonal neighbors which have already been identified as central tag voxels (this distinguishes them from tag voxels at the periphery which do have orthogonal neighbors classified as central tag voxels). The next step in the surface fitting is the specification of orthogonal masks (see FIG. 7). These are constructed using the per voxel derived surface normal $\mathbf{n_s}$ (see equation II-6). For each central tag voxel with voxel coordinates $\mathbf{p_v}$, the orthogonal mask coordinates $\mathbf{V_m}$ are defined as the collection of all voxels along the line (see arrows in FIG. 7):

$$\{(\mathbf{p_v} - \mathbf{n_s}h_m), (\mathbf{p_v} + \mathbf{n_s}h_m)\} \qquad \text{II-8}$$

The parameter $h_m$ is the mask height here set to $\frac{p_t}{4}$. For each tag voxel with orthogonal neighbors a tag surface point $\mathbf{p_s}$ can be derived using the orthogonal weighted average:

$$\mathbf{p_s} = \frac{\sum \mathbf{V_m W_m}}{\sum \mathbf{W_m}} \qquad \text{II-9}$$

where $\mathbf{W_m}$ are the orthogonal mask weights linearly derived from the image signal intensities. When applied to all voxels with orthogonal neighbors this produces a set of points describing the surface where each point location was determined depending only on its local orthogonal neighborhood. As such sharp transitions, gaps and shear interfaces are permissible and do not require special treatment. The next step is to assume a type of connectivity across these points to form the surface. This is done through a simple Delaunay triangulation whereby gaps and sheared interfaces are accounted for via removal of triangles with edge lengths longer than twice the largest voxel dimension of the image set (see gaps in FIG. 8). Finally to suppress the effects of noise and to reduce the stepped appearance induced due to the discrete nature of voxels, the surfaces are mildly smoothened using surface smoothening (HC-Laplacian smoothening[45] ensuring shape shrinkage is limited). The smoothening is based only on local connected neighborhoods (Laplacian umbrella's) and does not occur across gaps and sheared interfaces and thus such sharp transitions are appropriately maintained.

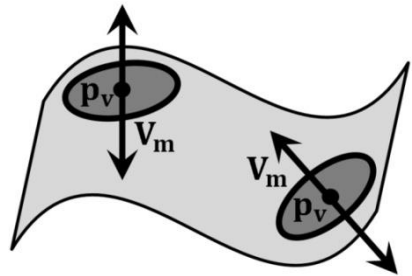

FIG. 7 Schematic visualisation of a tag surface, two tag voxel locations $\mathbf{p_v}$ and associated orthogonal masks $V_m$.

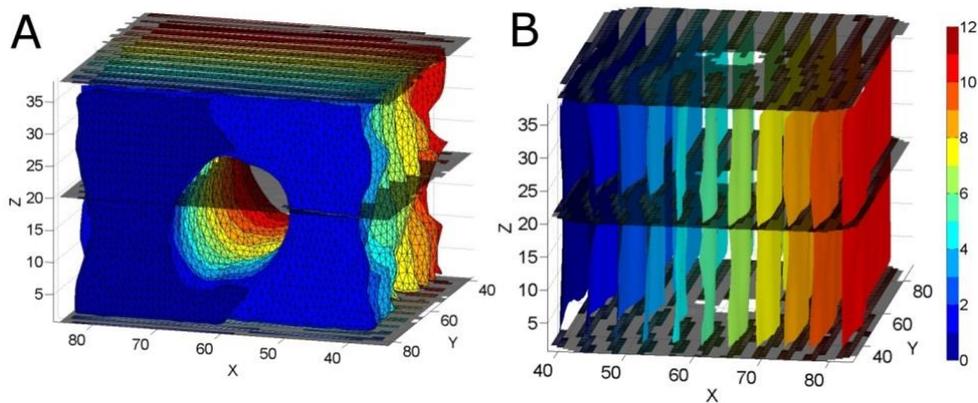

FIG. 8 Segmented surfaces for two of the phantom data sets. Note the triangulated appearance of the surfaces and how sheared interfaces and gaps e.g. at the core location are dealt with (A) and how curvature (due to indentation in Y direction) is captured (B). Axis units are in mm.

### II.D.4. Tag surface intersection determination

Since the prior analysis step produced triangulated surfaces a logical and simple method for obtaining tag surface intersections is to compute triangle intersections. Each intersection point is uniquely defined by the intersection of a triangle triplet. If each surface contains $n$ triangles then $n^3$ triplet combinations exist. Thus for computational efficiency the number of candidate triangles is first reduced. This is done by focusing the analysis only on those triangles that are closer than the longest occurring triangle edge length (based on nearest vertex search of Delaunay tessellation). For each surface this reduces the candidate triangles to only those located near a possible intersection point. The triangle intersection calculation is based on vector geometry and is detailed in the appendix. Since the tag feature segmentation provides numbered indices for each tag surface, each intersection point, being an intersection point of a tag surface triplet for a certain dynamic, is thus uniquely specified by a 4 digit tag-index $(T_1, T_2, T_3, d)$, where $d$ is the dynamic number and $T_i$ the tag numbers for each direction. Thus each tag point position **p** was stored in the tag point array $\mathbf{P}(T_1, T_2, T_3, d)$. This type of indexing avoids the need for tag intersection point tracking methods (e.g. point matching algorithms [46]).

### *II.D.5. Derivation of dynamic deformation*
In the current study deformation is measured following tracking of intersections of tag surfaces in 3 sets of mutually orthogonal tag surfaces (although other types of intersecting oblique orientations are permissible and would not require alteration of the methods presented here). Since field in-homogeneities may cause the initial tag shape to be non-planar displacement is defined with respect to tag points derived from tag-surfaces segmented for an initial configuration. The displacement array **U** is thus defined by:

$$\mathbf{U}(T_1, T_2, T_3, d_i) = \mathbf{P}(T_1, T_2, T_3, d_i) - \mathbf{P}(T_1, T_2, T_3, d_0) \qquad \text{II-10}$$

where $d_0$ is the appropriate reference or initial dynamic for the dynamic $d_i$. These per dynamic displacements are referred to as the individual displacement fields. The above is schematically illustrated in FIG. 9, as the tissue deforms the initial tags and intersections (light gray in FIG. 9A) obtain a new location (dark gray in FIG. 9A). This produces the first individual dynamic displacement field (green vectors in FIG. 9A). For each consecutive dynamic the deformed tissue is re-tagged (hence its initial coordinates may represent the same spatial coordinates but not the same tissue points as the previous dynamic) and undergoes an additional displacement (red vectors in FIG. 9B).

FIG. 9 C-D shows a schematic derivation of the so called cumulative dynamic displacement. The deformed state of the first (or prior) dynamic (light gray in FIG. 9C) is mapped into the displacement field of the second (or current) dynamic (transparent vectors in FIG. 9C). Through (natural neighbor and 3D Delaunay tessellation based) interpolation of the current displacement field onto the deformed state of the prior dynamic, it is possible to derive the displacement that the initial tissue points underwent during the second dynamic (dotted arrows in FIG. 9 C). The continuous mapping of the previous state into the current allows for the construction of a continuous cumulative displacement path over time (consecutive green and red arrows in FIG. 9D).

Each individual dynamic displacement measurement is derived following a single initial and a single deformed state hence possible displacement measurement errors are a function of two measurements. For the cumulative dynamic displacement however each displacement field is a function of all past measurements and thus measurement errors may propagate. A common approach is to apply regularization techniques and assumptions on the nature of the deformation and underlying constitutive properties. However implementation of such assumptions is not of interest to the current study. Therefore the approach presented here does not require *a priori* knowledge of the geometry and nature of the deformation and mechanical properties. Therefore cumulative displacement is derived using simple natural neighbor interpolation instead.

For the current study the cumulative displacement measures are required in order to derive accuracy measures (e.g. marker displacement prediction, see section II.E).

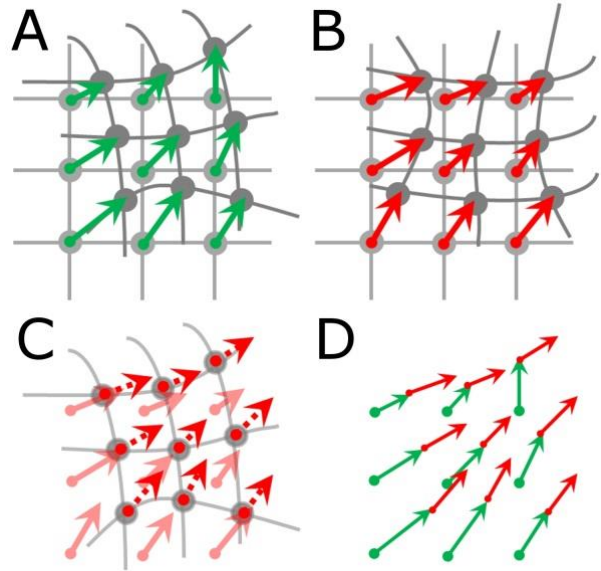

FIG. 9 Derivation of dynamic displacement (A-B). Derivation of cumulative dynamic displacement (C-D).

## II.E.  Analysis of precision and accuracy

As shown in Table I continuous dynamic tagging was applied in each direction during repeated indentations. Due to the repetitions for each direction a multitude of combinations can be made for computation of tag surface intersections and displacement fields. In order to limit computational time random data set combinations were chosen leading to 6 tag point location sets per dynamic allowing computation of 36 individual displacement fields per dynamic. Using the mean of all combinations of individual displacement fields a single mean cumulative displacement field was derived. The following precision and accuracy measures were evaluated:

*1) Precision of tag point location*

For each dynamic the difference of each tag point location combination ($n = 6$) with respect to the mean tag point location for that dynamic was calculated. This allowed computation and analysis of difference scatter clouds for each dynamic.

*2) Precision of individual dynamic displacement*

For each dynamic the deviation of displacement magnitude of each individual displacement field combination ($n = 36$) with respect to the average individual displacement magnitude was calculated.

*3) Accuracy of cumulative dynamic displacement*

Since deformation is tracked up to the end of the retraction phase of the motion cycle differences between the initial coordinates and the final locations of the total cumulative displacement are a measure of accuracy both in the phantom and *in-vivo*.

*4) Accuracy of displacement compared to marker tracking in the phantom.*

Using the mean cumulative displacement of the indentation phase of the motion cycle the marker locations in the deformed state can be predicted and compared to the real measured marker locations[44]. Comparison of true independently measured and predicted marker displacement marker thus yields a measure of the accuracy of the methods applied to the phantom.

Statistical analysis was performed using fitting of Gaussian models (see Matlab *gmdistribution* function and[47]) to the various difference measures. Each overall mean was defined as the root mean square (RMS) of the means in the X, Y and Z directions while the overall standard deviations are defined as the square root of the mean Eigen-value of the co-variance matrix. In addition RMS values were computed where appropriate allowing for comparison to values in the literature.

## III. RESULTS

## III.A. Precision of tag point location

For both the phantom and volunteer data 6 tag point location sets were derived for each dynamic and compared to the mean tag point locations for each dynamic. FIG. 10A-B below shows the difference scatter plots for all tag points (all combinations and for all dynamics) where is n=449495 and n=232391 for the phantom and volunteer data respectively.

For the phantom data the standard deviation was 44 µm (RMS of difference magnitudes 76 µm). The largest tag point location difference was 0.96 mm. Such outliers are however rare as differences larger than 0.28 mm were found in less than 1% of tag points. Similarly for the human data the standard deviation was 92 µm (RMS of difference magnitudes 160 µm) and the largest tag point location difference found was 1.73 mm. However, again such outliers are however rare since differences over 0.4 mm were found in less than 1% of tag points.

FIG. 10C-D demonstrates that no clear relationship between the standard deviations and the respective dynamic exists for either the phantom of the volunteer data.

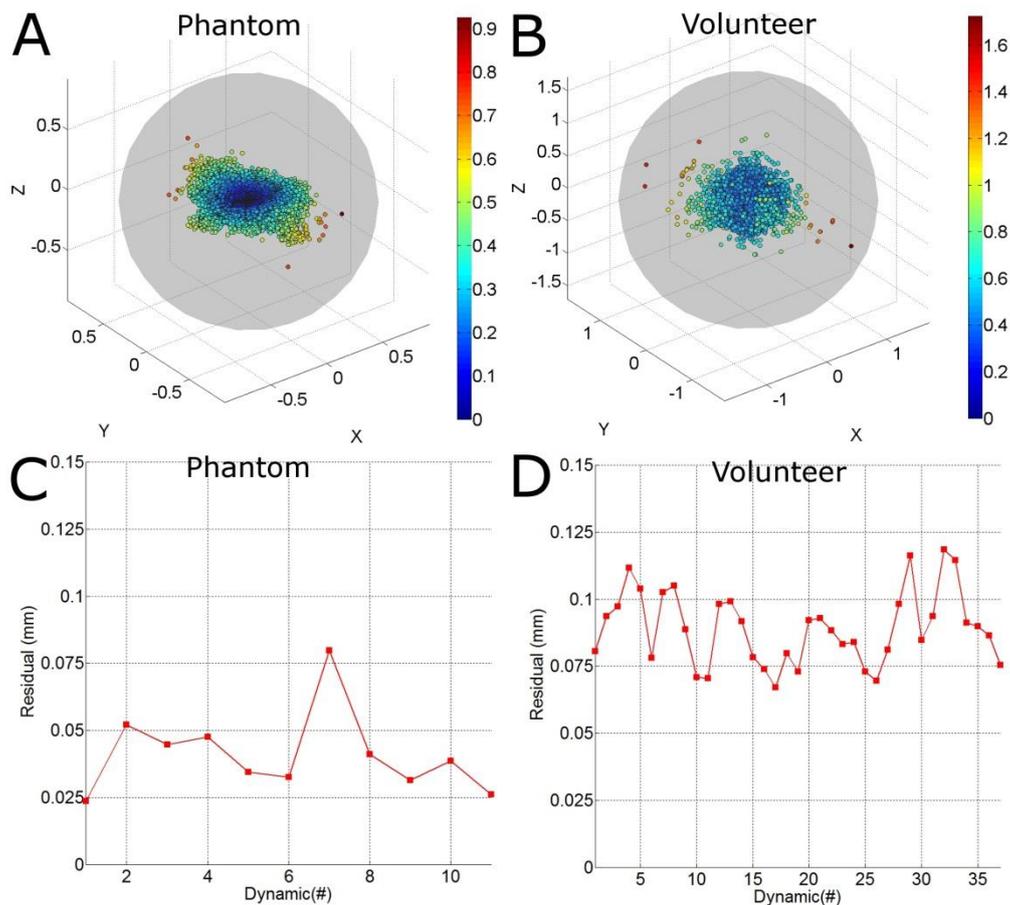

*FIG. 10 Tag point location difference scatter plots and circumspheres for all dynamics (A-B). Points shaded according to difference magnitude. In addition the means (dotted curve) and standard deviations (blocked curve) of the differences as a function of dynamic number (C-D). Images on the left are for the phantom and on the right are for the volunteer. All units are*

## III.B. Precision of individual dynamic displacement

For both the phantom and volunteer data a total of 36 individual displacement fields (FIG. 11) were derived for each dynamic and compared to the mean individual displacement for each dynamic. For the phantom data the overall standard deviation for the displacement magnitude differences was 61 µm (n=2567001). The largest difference found was 0.87 mm (outliers over 0.24 mm occurred in less than 1% of cases). The standard deviation was 91 µm for the volunteer data (n=1388343). The largest difference found was 1.44 mm (outliers over 0.29 mm occurred in less than 1% of cases).

Similar to the tag-point precision results no relationship across dynamics was observed. In addition as the scatter plots and distributions in FIG. 12 demonstrate, no relationship with displacement magnitude was observed as the standard deviation did not vary significantly with increasing displacements.

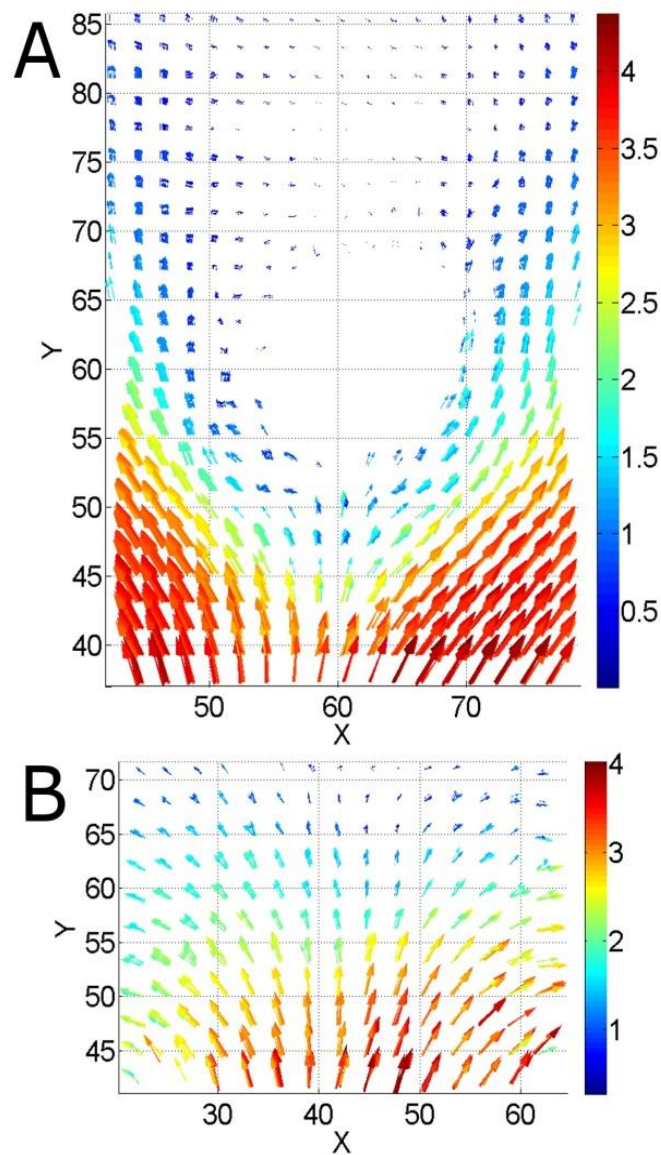

*FIG. 11 One of the individual displacement fields during the indentation phase for the phantom (A) and volunteer data (B). Displacement vectors are shown as arrows shaded towards magnitude. All units are in mm.*

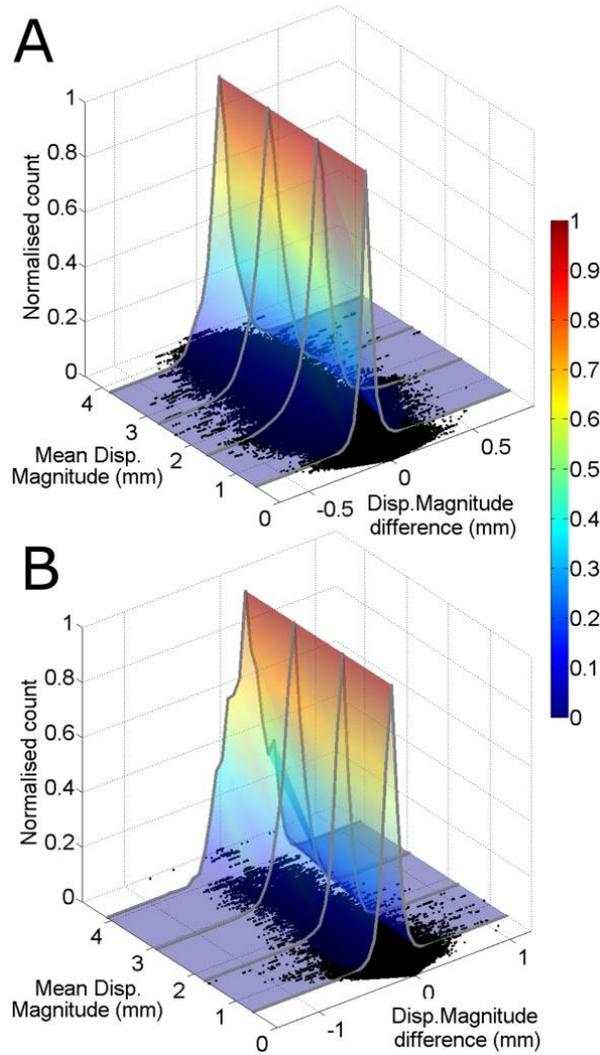

*FIG. 12 Normalized distributions (grey curves and shaded surface) of the displacement magnitude differences (for all combinations and all dynamics) as a function of the mean displacement magnitude for the phantom (A) and human volunteer data (B). The XY-planes of the graphs show scatter-plots (black dots) for all differences. All units are in mm.*

## III.C. Accuracy of cumulative dynamic displacement

Through analysis of the entire deformation cycle (indentation, hold and retraction phase) a complete cumulative displacement vector path could be reconstructed (FIG. 13) for both the phantom and volunteer data. The complex nature of the deformation induced by the indentation is evident from the curved motion paths observed. For both the phantom and volunteer data sets the differences between the start and end locations of the motion paths were derived in order to calculate measures of accuracy of the cumulative displacement measurement. FIG. 13D illustrates a selection of motions paths. For both the phantom and volunteer data it was found that differences were smallest for locations where displacement vectors maintain sufficient neighbors throughout all dynamics (such as region 1 in FIG. 13D) while largest errors were found at the periphery of the displacement field (e.g. region 2 FIG. 13D) where natural neighbor interpolation used for computation of the cumulative displacement is more limited. Displacement vectors at the periphery were therefore not included in the further analysis. As is evident from the difference scatter plots in FIG. 14 overall a good agreement was found with mean differences and standard deviations of 0.44 mm and 0.59 mm for the phantom (combination of 11 dynamics) and 0.40 mm and 0.73 mm for the human data (combination of 37 dynamics) respectively. As was mentioned in section II.D.5 constraint free derivation of cumulative displacement may be sensitive to error propagation. Hence some large differences were found with maxima of 2.8 mm and 3.5 mm respectively for the phantom and volunteer data. As shown in FIG. 15 no significant relationship with (cumulative) displacement magnitude was found and the mean of the differences did not vary significantly with displacement. However, for the phantom the mean of the difference was lower for displacement magnitudes under 5 mm but remain constant for larger displacements.

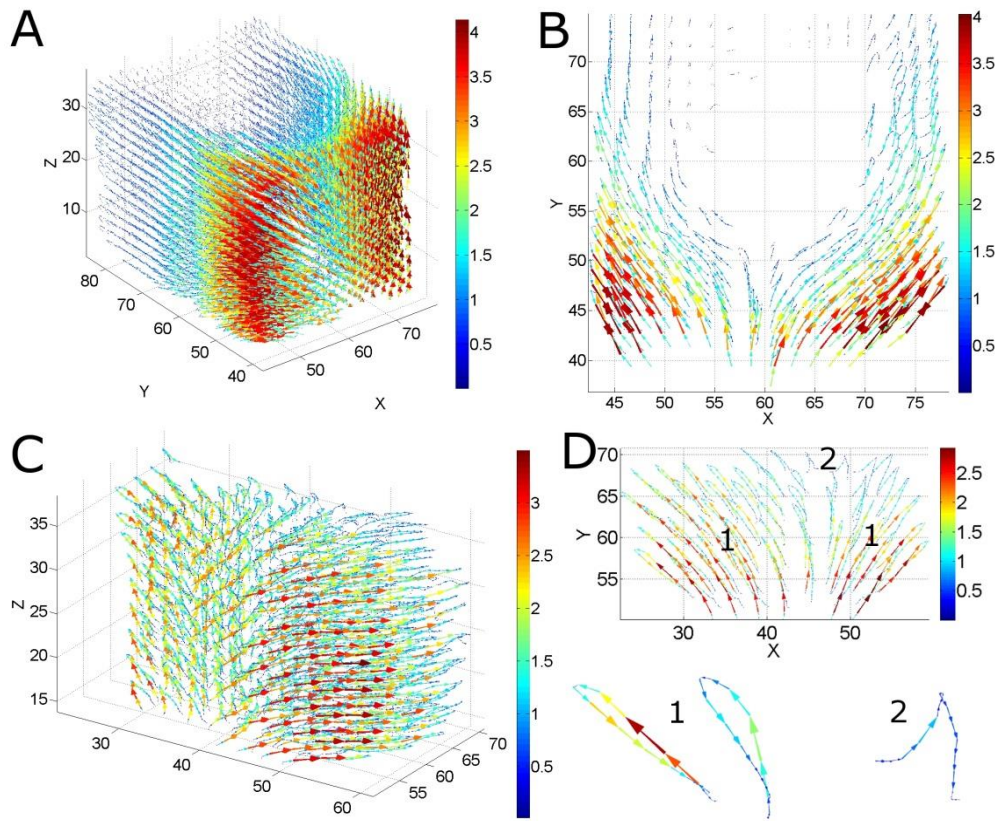

*FIG. 13 The cumulative displacement for the phantom (A-B) and volunteer data (C-D). Vector fields are shown and individual arrows shaded towards magnitude "Slice" views are shown in B and D. In addition a selection of motion paths are illustrated (bottom of D) to show results for locations embedded in (D-1) or on the edge of the displacement field (D-2). All units are in mm.*

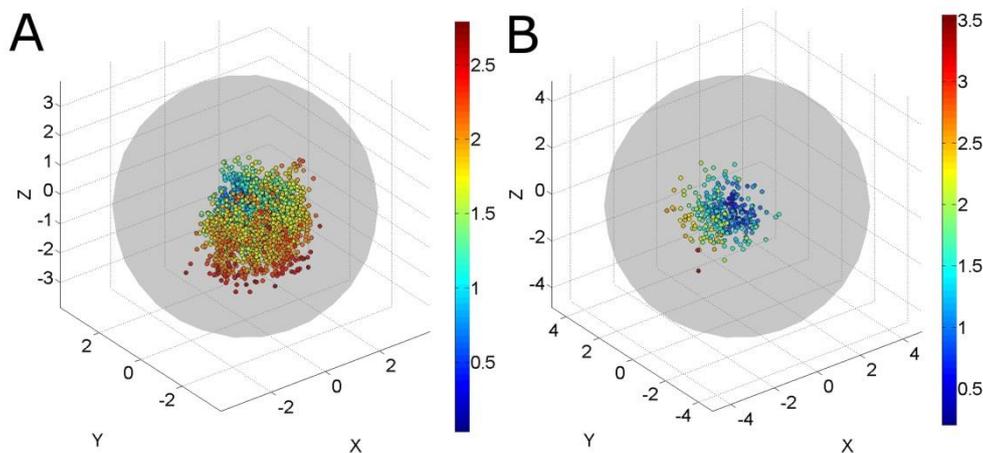

*FIG. 14 Scatter plots and their circumsphere for the cumulative displacement difference with respect to the initial for the phantom (A) and volunteer (B) data. Points are shaded according to difference magnitude. All*

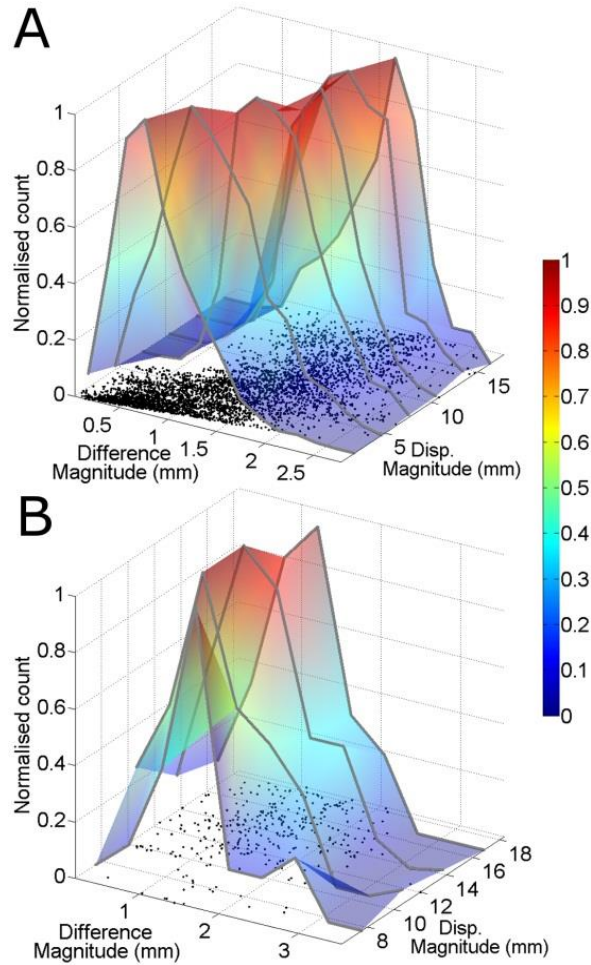

*FIG. 15 Normalized distributions (grey curves and shaded surface) of the location difference magnitudes as a function of the total cumulative displacement magnitude for the phantom (A) and human volunteer data (B). The XY-planes of the graphs show scatter-plots (black dots) for all differences. All units are in mm.*

## III.D. Accuracy of displacement in the phantom

FIG. 16A shows the average cumulative displacement for the indentation part of the motion cycle and marker locations (n=15) within the field of view for the initial and final deformed configuration. Using the average displacement field the marker locations in the deformed configuration were predicted (red in FIG. 16A) and could be compared to the true independently measured locations. The difference between the predicted marker displacement and the actual marker displacement provides a measure of the accuracy of the displacement measurement. The differences showed a mean of 0.35 mm and a standard deviation of 0.63 mm (X,Y and Z RMS values were 0.59, 1.02 and 0.41 mm respectively). The difference demonstrated no relationship with displacement magnitude. The largest difference found was 2.65 mm for a marker close to the edge of the displacement field, where the interpolation based computation of the cumulative displacement and marker prediction is based on a relatively limited number of points.

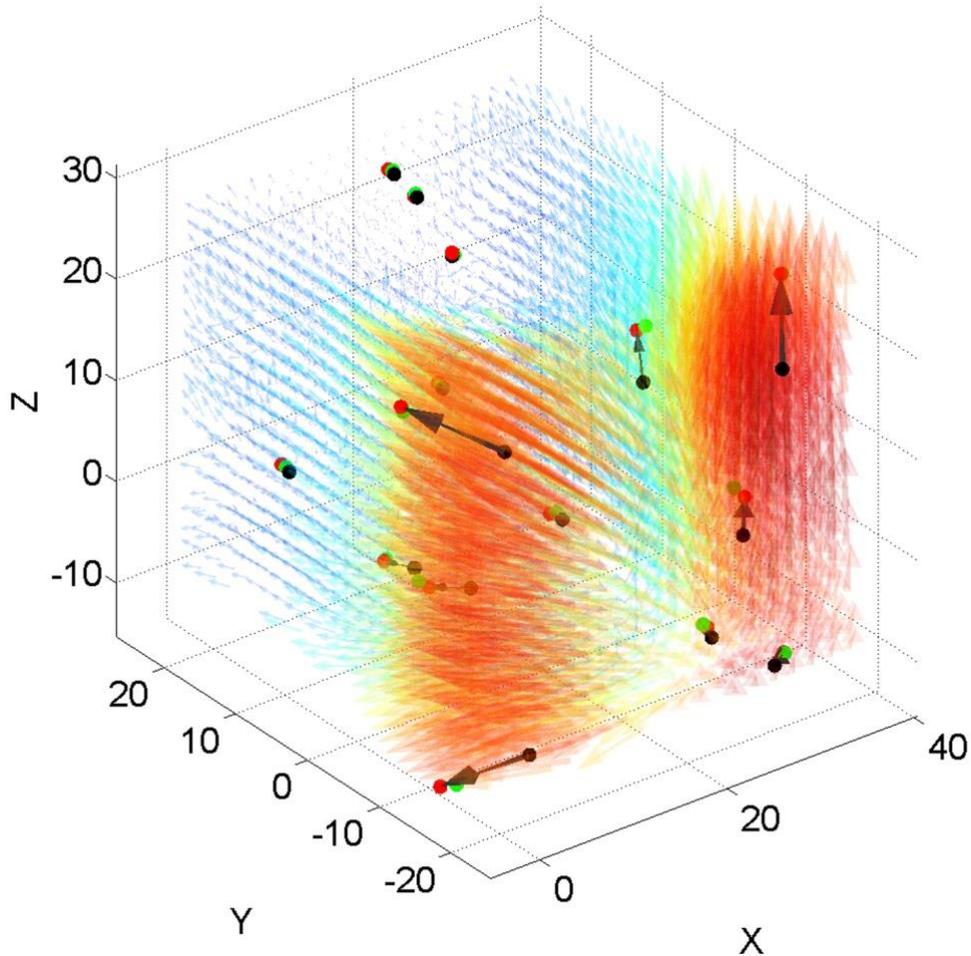

*FIG. 16 The cumulative displacement vector field for the phantom indentation phase shown as arrows shaded to magnitude (mm). Using the displacement field and the markers locations in the un-deformed configuration (black dots) the total marker displacement (black arrows and light gray dots) could be predicted and compared to the true final marker locations (dark gray dots). All units are in millimeter*

## IV. DISCUSSION

A novel SPAMM tagged MRI and fully automatic post-processing framework for the measurement of complex 3D dynamic soft tissue deformation following just 3 repeated motion cycles was presented. The techniques presented demonstrate dynamic measurement of complex 3D soft tissue deformation at sub-voxel accuracy and precision and were validated for 3.3-3.6Hz sampling of deformation speeds up to 12 mm/s.

A fully automatic post-processing framework is presented featuring the novel implementation of Gabor scale-space and filter bank assisted segmentation. In addition triangulated surfaces were fitted to the segmented tag extrema aided by Gabor filter bank derived estimates of local surface normal directions. Finally a dense grid (2x2x2 mm and 3x3x3 mm for the phantom and volunteer respectively) of tag points could be tracked following computation of tag surface triplet triangle intersections. Despite the fact that the post-processing methods are conceptually simple they are highly flexible as they enable tracking of tag features at any oblique orientation and undergoing complex deformations.

In the current study deformation is derived using 3 orthogonal SPAMM data sets and 3 repeated motion cycles only. This is a significant improvement over current methods involving many repetitions (e.g. in the order of 16 per slice [14]). Besides the reduction in scan time the presented methods have the potential to expand the application of SPAMM to the study of less periodic motions and motions which are difficult or uncomfortable to repeat. In this paper the techniques are applied for indentation based biomechanical tissue investigation where the minimization to 3 motion cycles ensured volunteer comfort and avoided repeatability issues associated with tissue preconditioning.

The soft tissue deformation measurement techniques were validated against marker tracking in a silicone gel phantom and evaluated *in-vivo* for the upper arm. Sub-voxel accuracy and precision levels were found. Below a comparison to literature is presented however, comparison is challenging since tissue sites, deformation modes, deformation speeds, and validation measures vary considerably. Recently Chen et al. 2010 [35] presented tagged MRI methods for cardiac deformation measurement whereby deformation was derived from the image data using Gabor filter banks, point matching and deformable models. Evaluation of their methods based a numerical phantom showed RMS displacement errors of 0.15~0.37 mm. Xu et al. 2010 [27] derived deformation from tagged MRI using optical flow methods and presented evaluation of the methods using simulated deformations showing X, Y and Z direction RMS errors of 0.43 mm, 0.45 mm and 1 mm respectively. In our previous work [17] where a similar validation set-up was used for static deformation measurement, comparison to marker tracking in a silicone gel phantom showed a mean difference of 72 μm and a standard deviation of 0.29 mm. In addition, displacement magnitude precision was analyzed and the standard deviation for displacement differences with respect to overall mean displacement was 72 μm for the phantom and 169 μm *in-vivo*. The accuracy and precision levels presented in the current study are of a similar magnitude to those in the literature. For instance for the current study differences between true and predicted marker locations showed a mean of 0.35 mm and a standard deviation of 0.63 mm.

Gabor filters have mainly been used for 2D and harmonic phase type analysis. In addition large standard deviations in the order of the tag period are commonly employed (e.g. [36, 39, 40]) which reduce the amount of local deformation information obtainable. In the current study 3D magnitude image data is analyzed using a combined 3D Gabor filter bank and scale-space enabling segmentation of tag features while deformation information is maximized using final standard deviations a third of the tag period in magnitude. The Gabor filters are also employed as "measurement tools" for the estimation of local tag feature orientation which could be used to aid surface fitting.

Since a major application of the work presented in this paper is the non-invasive analysis of tissue mechanical properties a constraint free (no assumptions on the nature of the mechanical properties or deformation) methodology for the computation of cumulative displacement was employed. Methods for the derivation of cumulative displacement from individual displacement fields often involve deformable models[35] and are akin to non-rigid registration methods (e.g. related to deformable image registration and finite element analysis[48]). However these approaches require *a priori* knowledge and assumptions on the nature of the deformation and/or the mechanical properties of the tissue and were thus not of interest to the current study. In cases where the mechanical behavior and deformation of tissue is well understood the implementation of these methods (combined with the individual displacement measurements presented) may provide an improvement on the accuracy achieved with the constraint free derivation of cumulative displacement presented here.

Some limitations need to be addressed. Due to the dynamic nature of the methods presented here motion occurring during the read-out is inevitable. This has a temporal averaging effect on the appearance of the tag features and as such may lead to underestimation of deformation. This effect is however deemed small especially given the presented results.

The current study presents the validation of the measurement of complex dynamic 3D soft tissue deformation for ~3.3Hz sampling of a ~20 mm deep indentation (and retraction) at speeds of 12 mm/s leading to individual dynamic displacement magnitudes of up to 4 mm (approximately 1 tag period). Although the validation of higher speeds is not presented here the techniques are not limited to deformation speeds of 12 mm/s as the delay (which is user defined) and read-out (which depends on resolution but also scanner hardware) times can be adjusted for higher speeds e.g. to obtain similar displacements magnitudes in individual dynamics. Hence the scanning protocol can be customized to the expected deformation speeds and magnitudes.

The response of a Gabor filter is most appropriate if the image features locally resemble its appearance. Severe local deformations within regions the size of the Gaussian envelope may cause the image features to depart from the Gabor filter appearance. However as mentioned before tag feature deformations can be constrained (here to <tag period) through appropriate high speed imaging with respect to the tissue deformation.

Due to the current musculoskeletal application a 2-channel flexible coil system was implemented. In addition analysis of the upper arm for supine volunteers resulted in suboptimal positioning of the region of interest at the edge of the bore. Other applications featuring more advanced coil types and more central positioning may thus yield better results.

The techniques presented here were developed for non-clinical (e.g. biomechanical research) applications. Future work may allow these methods to be applied in a clinical setting. Nevertheless in clinical applications signal to noise conditions may vary and thus further evaluation may be required. However this study has demonstrated the techniques to be robust under relatively poor imaging conditions i.e. due to the single-shot nature of the acquisitions, the use of the 2-channel flex coils, and positioning of the field of view within the bore. This lead to relatively low signal to noise levels even in comparison to typical clinical cardiac SPAMM tagging for which repeated acquisitions and >6-channel coils are common.

Currently 3 motion cycles are required for the computation of 3D deformation. Thus the methods presented are limited to the analysis of motion types which allow such repeatability reliably. For the application of computer controlled indentation presented such motions are easily and reliably repeated and synchronized using motion triggering. However un-triggered motion analysis of repeated motions can also be facilitated since temporal synchronization of the 3 SPAMM directions can be achieved in post processing. Ideally however 3D deformation should be derivable from unrepeated motion. This would enable the imaging of non-periodic arbitrary motion (e.g. bowel motion[49]). Although non-segmented acquisitions of grid tagged volumes (simultaneous tag modulation tagging in 3 mutually orthogonal directions in a single image volume) is possible, the triple saturation pattern significantly reduces signal intensities hindering analysis of deformation at present.

Future work will focus on the combination of the current techniques with inverse finite element analysis for the analysis of the mechanical properties of human skeletal muscle tissue.

# V. CONCLUSIONS

Novel SPAMM tagged MRI based methods are presented for high the measurement of complex dynamic 3D soft tissue deformation following just 3 motion cycles. Deformation is derived using a novel and fully automatic Gabor scale-space and filter bank based post-processing framework. The techniques were validated using marker tracking in a silicone gel soft tissue phantom for indentation induced dynamic deformation measurement. In addition *in-vivo* evaluation for the measurement of indentation induced tissue deformation of the biceps region of the upper arm was performed. The techniques presented demonstrate dynamic measurement of complex 3D soft tissue deformation at sub-voxel accuracy and precision and were validated for 3.3-3.6Hz sampling of deformation speeds up to 12mm/s. As only 3 deformation cycles are required the techniques presented are to the authors' knowledge the fastest currently available for the derivation of 3D dynamic deformation.

## ACKNOWLEDGMENTS

This research was partly funded by Science Foundation Ireland (Research Frontiers Grant 06/RF/ENMO76). The financial support agencies were not involved in designing and conducting this study, did not have access to the data, and were not involved in data analysis and/or preparation of this manuscript.

# APPENDIX: INTERSECTION POINT DETERMINATION

The intersection point **p** for 3 planes is defined by:

$$\mathbf{p} = \frac{(\mathbf{x}_1 \cdot \mathbf{n}_1)(\mathbf{n}_2 \times \mathbf{n}_3) + (\mathbf{x}_2 \cdot \mathbf{n}_2)(\mathbf{n}_3 \times \mathbf{n}_1) + (\mathbf{x}_3 \cdot \mathbf{n}_3)(\mathbf{n}_1 \times \mathbf{n}_2)}{\det([\mathbf{x}_1 \quad \mathbf{x}_2 \quad \mathbf{x}_3])} \qquad 0\text{-}1$$

where $\mathbf{x}_1$, $\mathbf{x}_2$ and $\mathbf{x}_3$ are arbitrary points on each plane and $\mathbf{n}_1$, $\mathbf{n}_2$ and $\mathbf{n}_3$ are the face-normals. Next the validity of the intersection point (if existent) is determined by checking whether it is found either inside and/or on each of the triangle faces using:

$$L_{on} = \begin{cases} 1 & \left(\sum c_1 = 0 \vee \sum c_2 = 0 \vee \sum c_3 = 0\right) \wedge \left(\sum c_1 \geq 0 \vee \sum c_2 \geq 0 \vee \sum c_3 \geq 0\right) \\ 0 & \end{cases}$$

$$\mathbf{c}_1 = (\mathbf{v}_2 - \mathbf{v}_1) \times (\mathbf{p} - \mathbf{v}_1)$$
$$\mathbf{c}_2 = (\mathbf{v}_2 - \mathbf{v}_3) \times (\mathbf{p} - \mathbf{v}_2)$$
$$\mathbf{c}_3 = (\mathbf{v}_3 - \mathbf{v}_1) \times (\mathbf{p} - \mathbf{v}_3)$$

$$L_{in} = \begin{cases} 1 & d_1 = 1 \wedge d_2 = 1 \wedge d_3 = 1 \\ 0 & \end{cases} \qquad 0\text{-}2$$

$$d_1 = \mathbf{c}_1 \cdot \mathbf{n}$$
$$d_2 = \mathbf{c}_2 \cdot \mathbf{n}$$
$$d_3 = \mathbf{c}_3 \cdot \mathbf{n}$$

$$L_v = \begin{cases} 1 & L_{in} = 1 \vee L_{on} = 1 \\ 0 & \end{cases}$$

Here $\mathbf{v}_i$ represent the triangle vertices and **n** its face-normal. An intersection point **p** is valid ($L_v = 1$) if it is inside ($L_{in} = 1$) or on ($L_{on=1}$) each of the triangles.